\documentclass[10pt,twocolumn]{article}
\usepackage[a4paper,margin=0.72in,columnsep=0.28in]{geometry}
\usepackage{mathptmx}
\usepackage[T1]{fontenc}
\usepackage{microtype}
\usepackage{amsmath,amssymb,amsthm}
\usepackage{booktabs}
\usepackage{graphicx}
\usepackage{tikz}
\usetikzlibrary{positioning,arrows.meta,shapes.geometric}
\usepackage{algorithm}
\usepackage{algpseudocode}
\usepackage{enumitem}
\setlist{nosep,leftmargin=1.2em}
\usepackage[hidelinks]{hyperref}
\usepackage{caption}
\usepackage{titlesec}
\titlespacing*{\section}{0pt}{8pt}{3pt}
\titlespacing*{\subsection}{0pt}{6pt}{2pt}
\titleformat{\section}{\normalfont\large\bfseries}{\thesection}{0.5em}{}
\titleformat{\subsection}{\normalfont\normalsize\bfseries}{\thesubsection}{0.5em}{}
\newtheorem{proposition}{Proposition}

\theoremstyle{remark}

\newcommand{\bundle}{\mathbf{a}}
\newcommand{\rec}{r}
\newcommand{\LCNB}{\textsc{LCNB}}

\title{\vspace{-0.6em}\textbf{Regime-Uncertain Data Transactions:\\ A Game-Theoretic Framework for Self-Adapting and Self-Diagnosing Multi-ML-Agent Systems}\vspace{-0.2em}}
\author{Chethana Prasad Kabgere\\
\small  \\
\small \textit{Draft for workshop submission. Code: \texttt{eregime} (released with this paper).}}
\date{}

\begin{document}
\maketitle
\vspace{-1.2em}

\begin{abstract}
\noindent Autonomous ML agents in large organisations exchange data and model outputs across organisational and legal boundaries.
Which rules bind a given exchange is not fixed. It depends on where the data subjects live, where the agents run, who owns them,
whom they target, and which version of each rule is in force. Tags that describe the data can be noisy, and a counterparty can overclaim its safeguards.
We model one exchange as a \emph{regime-uncertain transfer game} between two ML agents, with an enforcer as a third player.
Rules enter as hard constraints with a strictest-wins structure. Agents hold a Bayesian belief over which rules apply, can pay to unlock safeguards or verify claims, and then bargain over the remaining legal bundles.
Our method, Legally-Constrained Nash Bargaining (\LCNB), implements this with a Monotonic Concession Protocol using Zeuthen's rule.
We prove four properties of the model and test it in a synthetic simulator with GDPR, DPDP, CCPA/CPRA and a stylised stricter regime.
In two of three stress cases where residence and citizenship differ, a citizenship-based rule violates in 21--37\% of shared records, and \LCNB{} in under 1\%.
The results also show limits. With our default economics, \LCNB{} shares almost only aggregates, and its low violation rate follows partly from that.
Violation tolerance, tag noise and enforcement strength matter more than the choice of bargaining solution.
\end{abstract}

\section{Introduction}
Large organisations run many ML agents. A demand-forecasting agent in one company may need a supplier-cost estimate from an agent in another.
Each exchange is a data transaction. Many things decide whether the transaction may happen and in what form: ownership, purpose, internal policy, incentives, trust and law.
This paper studies one of them, and treats it as one part of a larger transaction problem.
The part is the following. \emph{The set of rules that applies to a transaction is itself uncertain.}

Consider a record of a person who lives in one region, held by an agent that runs in a second region, owned by a company in a third, and sent to an agent in a fourth.
A rule may apply because of where the person lives, where the processing happens, where the owner is established, or whom the owner targets.
A person's citizenship is often not an input at all.
Rules also change. India's DPDP Act, for example, has operative provisions that commence on a fixed future date \cite{dpdp2023}.
In practice an agent sees tags that may be missing or wrong, and a counterparty's statement of its own safeguards that may be wrong too.

We ask three questions. First, how should an agent \emph{adapt}: which rules apply to this transfer?
Second, how should it \emph{diagnose}: what to do when tags are noisy, rules overlap or conflict, versions change, or a claim cannot be trusted?
Third, how should two agents \emph{agree} on the form of the transfer: transformation, legal basis, safeguards, retention and purpose limits?

\paragraph{Contributions.}
\begin{enumerate}
\item A formal model, the \emph{regime-uncertain transfer game} (Sections~\ref{sec:model}--\ref{sec:game}). Rules are constraints over an ordinal bundle space. The strictest rule per dimension binds.
\item An algorithm, \LCNB{} (Section~\ref{sec:lcnb}). It combines a Bayesian belief over applicable rules, an unlock search, costly verification, and Nash bargaining realised by the Monotonic Concession Protocol.
\item Four analytical results (Section~\ref{sec:analysis}): monotone feasibility, the cost of tag noise, an enforcement threshold, and the cost of stale rule versions.
\item A reproducible simulator and eleven experiments (Sections~\ref{sec:setup}--\ref{sec:results}). We report where \LCNB{} helps, where it does not, and where an expected component did not fire.
\end{enumerate}

\paragraph{Scope.}
The legal rules are stylised constraints taken from public texts. They are not legal advice, and the paper makes no claim about legal compliance in practice.
The technical aim is to show how agents can reason about overlapping, uncertain constraints and still agree on a transfer.

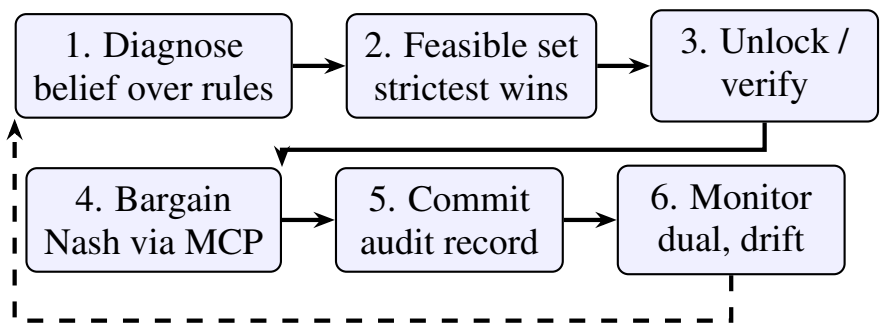
\begin{figure}[t]
\centering
\begin{tikzpicture}[
  node distance=0.32cm and 0.35cm,
  box/.style={draw, rounded corners=2pt, minimum height=0.62cm, minimum width=1.5cm, font=\scriptsize, align=center, fill=blue!6},
  arr/.style={-{Stealth[length=4pt]}, thick}]
  \node[box] (d) {1. Diagnose\\belief over rules};
  \node[box, right=of d] (f) {2. Feasible set\\strictest wins};
  \node[box, right=of f] (u) {3. Unlock /\\verify};
  \node[box, below=of d] (b) {4. Bargain\\Nash via MCP};
  \node[box, right=of b] (c) {5. Commit\\audit record};
  \node[box, right=of c] (m) {6. Monitor\\dual, drift};
  \draw[arr] (d)--(f); \draw[arr] (f)--(u); \draw[arr] (u.south) -- ++(0,-0.18) -| (b.north east);
  \draw[arr] (b)--(c); \draw[arr] (c)--(m);
  \draw[arr, dashed] (m.south) -- ++(0,-0.3) -| (d.south west);
\end{tikzpicture}
\caption{\LCNB{} pipeline for one transaction. The dashed arrow shows re-diagnosis when tags, leakage budget or rule versions change.}
\label{fig:pipeline}
\end{figure}

\section{Related Work}
\textbf{Normative multi-agent systems} study how norms constrain agents \cite{boella2006}. Most work is symbolic. We use a numeric constraint structure so that rules can enter utilities, beliefs and bargaining.
\textbf{Privacy-preserving learning} offers differential privacy \cite{dwork2014} and federated learning \cite{mcmahan2017}. These control what a transfer reveals. They do not decide which rules apply to the transfer.
Re-identification risk motivates our empirical leakage check \cite{sweeney2002,narayanan2008}.
\textbf{Automated negotiation} gives bargaining protocols for agents. We use Nash's solution \cite{nash1950}, Kalai and Smorodinsky's \cite{kalai1975}, Rubinstein's alternating offers \cite{rubinstein1982} and the Monotonic Concession Protocol with Zeuthen's rule \cite{zeuthen1930,rosenschein1994}.
The Myerson--Satterthwaite theorem limits what bilateral bargaining can achieve under private values \cite{myerson1983}.
\textbf{Games with incomplete information} \cite{harsanyi1967}, \textbf{costly state verification} \cite{townsend1979}, \textbf{Stackelberg and security games} \cite{basar1999,tambe2011} and \textbf{constrained Markov decision processes} \cite{altman1999,littman1994} supply the other parts.
We combine them for one problem: legally uncertain data transactions between ML agents.

\section{Model}\label{sec:model}
\subsection{Records, parties and bundles}
A \emph{record} has attributes $\rec=(\omega,s,c,o,h,\ell)$: residence region $\omega\in\{\text{EU, CA, IN, CN, OTHER}\}$, sensitive flag $s$, consent evidence $c\in\{\text{none, general, explicit}\}$, opt-out $o$, child flag $h$ and legitimate-use flag $\ell$.
There are $|\mathcal R|=240$ attribute combinations. Citizenship is carried as a control feature. No rule reads it.
A \emph{party type} is $\tau=(\text{host},\text{establishment},\text{role},\text{targets})$, where role is processor, third party or controller.

A \emph{bundle} $\bundle=(T,B,M,R,P)$ has five ordinal dimensions, and a higher level is more protective:
transformation $T$ (raw, pseudonymised, anonymised view, aggregate), legal basis $B$ (five levels up to explicit or parental consent),
transfer mechanism $M$ (none, processor contract, standard clauses, security assessment), retention $R$ and purpose binding $P$.
There are $|\mathcal A|=960$ bundles. Not every bundle can be held: the \emph{effective bundle}
$\bar\bundle(\bundle,\rec)=(T,\min(B,\text{cap}_B(c)),\min(M,\kappa_M),R,P)$ caps the basis by the consent evidence on file and the mechanism by what the pair can actually set up ($\kappa_M$).

\subsection{Rules as constraint compilers}
Each rule $k\in\{\text{GDPR, DPDP, CCPA/CPRA, PIPL-like}\}$ has a \emph{trigger} $\chi_k(\rec,\tau_S,\tau_R,v)\in\{0,1\}$ and a \emph{requirement} $\rho_k(\rec,\tau_S,\tau_R,v)\in\mathbb N^5\cup\{\text{BLOCK}\}$, both for rule version $v$.
Triggers use residence, establishment, processing location, targeting and commencement. Requirements are minimum levels per dimension.
A transformation level $T\ge 3$ is treated as anonymous and exempt. Bundle $\bundle$ \emph{violates} rule $k$ on record $\rec$ if
\begin{equation}
V_k(\bundle,\rec)=\chi_k\cdot\mathbb 1[T<3]\cdot\mathbb 1[\bar\bundle(\bundle,\rec)\not\ge\rho_k]=1 .
\end{equation}
Under overlap the \emph{strictest rule per dimension} binds: $\rho^*(\rec)=\bigsqcup_{k:\chi_k=1}\rho_k(\rec)$, the component-wise maximum (a lattice join).
The \emph{lawful set} is $F(\rec)=\{\bundle:\forall k,\ V_k(\bundle,\rec)=0\}$.
Appendix~A lists the clauses behind each rule and marks which we checked in public sources.

\subsection{Place in the transaction lifecycle}
A data transaction is initiated, negotiated, authorised, executed, monitored and adapted. Regime uncertainty enters at negotiation and authorisation (which bundles are admissible), and again at monitoring and adaptation (tags and versions change).
Ownership, purpose, internal policy, incentives and trust enter our model as utility terms, the purpose and retention dimensions, and a counterparty reputation. We do not model them in depth.

\section{The Regime-Uncertain Transfer Game}\label{sec:game}
\paragraph{Players and nature.}
The sender $S$, the receiver $R$ and an enforcer $E$ play. Nature draws a transaction: the records, the receiver's role, the true rule versions, the pair's true mechanism capacity $\kappa_M$ and whether the receiver overclaims.

\paragraph{Information.}
Agents see noisy tags $\tilde\rec$. Each feature is missing with probability $p_m$ or wrong with probability $p_w$.
The receiver's policy card states a capacity $\hat\kappa_M\ge\kappa_M$ (equal if honest, one level higher if it overclaims).
Agents may hold a stale view of the rule versions.

\paragraph{Belief (self-diagnosis).}
For a group $g$ of records with the same observed tags, the posterior over attributes factorises over features:
$b_g(\rec)\propto\prod_f\pi_f(\rec_f)\,L(\tilde\rec_f\mid\rec_f)$.
Because $|\mathcal R|=240$ we compute it exactly. The belief that rule $k$ applies is $P_g(k)=\sum_\rec b_g(\rec)\chi_k(\rec)$.
An action is \emph{admissible} if its violation probability is at most a tolerance $\epsilon$:
\begin{equation}
\mathcal A_\epsilon(b_g)=\Big\{\bundle:\ b_g\big(\{\rec:\exists k,\ V_k(\bundle,\rec)=1\}\big)\le\epsilon\Big\}.
\end{equation}

\paragraph{Actions.}
$S$ may verify the receiver's claim at cost $c_v$, request consent (raising the basis cap to explicit) at cost $c_B$ per record, negotiate a security assessment (raising $\kappa_M$ to its maximum) at a fixed cost, propose a bundle, or refuse. Each unlock succeeds with a probability $q_u$.
$E$ commits to an audit rate $p_a$ before play. Agents observe it.

\paragraph{Utilities.}
For a group of $n$ records and bundle $\bundle$, with belief $b$,
\begin{align}
u_S&=n\,v_S\,I(T)-n\,c_B(B)-\tfrac12 c_M(M)\notag\\
&\quad-\lambda_S L-\sigma\,\Pi-\tfrac12\mu L,\\
u_R&=n\,v_R\,I(T)\rho_R(R)\rho_P(P)-\tfrac12 c_M(M)\notag\\
&\quad-\lambda_R L-(1-\sigma)\Pi-\tfrac12\mu L.
\end{align}
$I(T)$ is the information value of the transformation. $L$ is leakage, $\Pi=p_a\,n\,\mathbb E_b[\kappa(\rec)\sum_k f_kV_k]$ is the expected penalty with fines $f_k$ and a sensitivity multiplier $\kappa$.
$\lambda_S,\lambda_R$ are the owners' leakage weights, $\sigma$ is the sender's share of penalties, and $\mu$ is a dual variable for a leakage budget.
The disagreement point is $d=(0,0)$, meaning no transfer.

\paragraph{Solution concept.}
We do not claim that \LCNB{} is an equilibrium of the full Bayesian game. It is a \emph{mediated} solution. Given beliefs, a mediator that knows both utilities computes the bargaining outcome inside the admissible set.
Zeuthen risks use only each agent's own utility and the opponent's public offer, but the Pareto frontier needs a mediator (Section~\ref{sec:lcnb}).

\section{The \LCNB{} Algorithm}\label{sec:lcnb}
\begin{algorithm}[t]
\caption{\LCNB{} for one transaction}\label{alg:lcnb}
\begin{algorithmic}[1]
\small
\State Group records by observed tags; compute posteriors $b_g$ with versions $\hat v$ (lagged).
\State \textbf{Verify:} if $\hat p_{\text{lie}}\cdot\Delta>c_v$, learn $\kappa_M$; update the Beta reputation.
\For{each group $g$}
  \State $\mathcal I\gets\{\bundle\in\mathcal A_\epsilon(b_g):u_S\ge0,u_R\ge0\}$
  \State $\bundle^\star\gets$ Nash solution on $\mathcal I$ via MCP--Zeuthen
  \For{each unlock $u$}
    \State $G_u\gets q_u\,(W(\bundle^\star_u)-W(\bundle^\star))-\kappa_u$
  \EndFor
  \If{$\max_u G_u>0$}\;attempt it; on success use $\bundle^\star_u$
  \EndIf
  \State Execute; log the offer and evidence.
\EndFor
\State Update the leakage dual: $\mu\gets\max(0,\mu+\eta(L_t-\bar b))$.
\end{algorithmic}
\end{algorithm}

\textbf{Bargaining.} \LCNB{} maximises the Nash product $(u_S-d_S)(u_R-d_R)$ over the individually rational admissible set.
It reaches the Nash point with the Monotonic Concession Protocol. Each agent opens with its best Pareto offer.
Let $\text{risk}_i=(u_i(\text{own})-u_i(\text{other}))/u_i(\text{own})$. The agent with the smaller risk concedes the least amount that makes its opponent's risk smaller than its own. The protocol stops when the offers cross.
Our implementation matches the Nash point on a 400-point frontier test (\texttt{tests/test\_negotiation.py}).

\textbf{Unlock.} Consent and mechanism unlocks have costs $\kappa_u$ and success probabilities $q_u$. The rule in lines 5--9 attempts the one with the largest positive expected gain, and falls back to $\bundle^\star$ on failure.

\textbf{Verification.} $\Delta$ is the expected penalty increase if the receiver's claimed capacity is one level too high. $\hat p_{\text{lie}}$ is the posterior mean of a Beta distribution updated by past verifications. This is a value-of-information rule in the spirit of costly state verification \cite{townsend1979}.

\textbf{Repeated play.} A leakage budget $\bar b$ per tick is enforced by dual ascent on $\mu$, a Lagrangian treatment of a constrained Markov game \cite{altman1999}.

\section{Analysis}\label{sec:analysis}
\begin{proposition}[Monotone feasibility]\label{prop:mono}
If $\bundle\le\bundle'$ component-wise and $\bundle\in F(\rec)$, then $\bundle'\in F(\rec)$. Also $F(\rec)\supseteq\{\bundle:T\ge3\}\ne\emptyset$, and $F(\rec)=\{T\ge3\}\cup\{\bundle:\bar\bundle(\bundle,\rec)\ge\rho^*(\rec)\}$.
\end{proposition}
\begin{proof}
The effective bundle is monotone in $\bundle$. Each requirement is a per-dimension threshold, so joint compliance with all active rules equals compliance with their join $\rho^*$.
The exemption $T\ge3$ is also upward closed.
\end{proof}
Monotonicity means that adding protection never hurts legality, and that the feasible set is described by one minimal requirement vector per record. Strictest-wins is exact when rules are thresholds.
Real rules can also contain prohibitions on combinations. We map these to BLOCK or to an empty set outside the exemption.

\begin{proposition}[Cost of tag noise]\label{prop:noise}
Let $\bundle$ have $T<3$ and let $f$ be a tag that is missing for a record. If some value $x$ of $f$ switches on a rule whose requirement exceeds $\bar\bundle$, and the other features are independent, then $b_g(\{V=1\})\ge\pi_f(x)$.
So $\bundle$ is admissible only if $\pi_f(x)\le\epsilon$. If $\epsilon<\min_x\pi_f(x)$ over such values, every admissible bundle for that record is exempt.
\end{proposition}
\begin{proof}
With the tag missing, the posterior of $f$ equals its prior. The violating set contains all attribute combinations with $f=x$, which has probability $\pi_f(x)$.
\end{proof}
In our default setting the EU residence prior is 0.30, and $\epsilon=0.02$. A missing residence tag therefore forces aggregate sharing for that group.
A wrong tag is milder. With $p_w=0.05$ the posterior puts mass about $0.0125$ on each other region, which is below $\epsilon$.

\begin{proposition}[Enforcement threshold]\label{prop:enf}
Take a lawful bundle $\bundle_L$ and an unlawful bundle $\bundle_U$ that violates rules $K_U$ on a record. Let $\Delta W=W_0(\bundle_U)-W_0(\bundle_L)>0$ be the joint payoff gain from violating, before penalties.
A planner who maximises $W=u_S+u_R$ prefers $\bundle_L$ if and only if
\begin{equation}
p_a>p^*=\frac{\Delta W}{\kappa(\rec)\sum_{k\in K_U}f_k}.
\end{equation}
\end{proposition}
\begin{proof}
Expected penalty is $p_a\kappa\sum_{k\in K_U}f_k$ per record. Compare payoffs. The penalty shares $\sigma$ and $1-\sigma$ sum to one in $W$.
\end{proof}
The threshold falls as fines and sensitivity rise, and it rises with the gain from violating. Welfare in our simulator excludes reputation and other costs of violating, so $p^*$ is a lower bound on a real deterrence threshold under that model.

\begin{proposition}[Cost of stale versions]\label{prop:drift}
Suppose a rule commences at $t_0$ and an agent re-reads versions with lag $\Lambda$. Let $q$ be the probability that a bundle chosen under the stale belief violates the new rule. Then the expected number of extra violating records per transaction is about $q\,n$ for $t\in[t_0,t_0+\Lambda)$ without re-detection beyond that window it is zero, while an agent that never re-reads keeps rate $q$ for all $t\ge t_0$.
\end{proposition}
\noindent The statement is an accounting identity once $q$ is fixed. The value of $q$ is empirical (Section~\ref{sec:drift}).

\begin{proposition}[Nash and concessions]\label{prop:nash}
On a finite Pareto frontier, the Monotonic Concession Protocol with Zeuthen's rule ends at the Nash bargaining solution (up to discretisation).
\end{proposition}
\noindent This is a known result \cite{rosenschein1994}. We do not reprove it. We check it numerically.
By the Myerson--Satterthwaite theorem \cite{myerson1983} no mechanism makes bilateral trade ex-post efficient under private values and individual rationality, so we measure the loss against an oracle instead of assuming none.

\section{Experimental Setup}\label{sec:setup}
\paragraph{Simulator.}
Two organisations $A$ (sender) and $B$ (receiver) exchange batches of 10 synthetic records per tick. Records carry the true attributes of Section~\ref{sec:model} and noisy observed tags.
The rule library encodes GDPR, DPDP, CCPA/CPRA and a stylised stricter regime with versions (Appendix~A). A ground-truth oracle computes the true lawful set, using true tags, versions and capacities.
All methods see the same transaction stream (common random numbers).

\paragraph{Scenarios.}
Table~\ref{tab:scn} lists them. The first five are natural cases. The last three are stress tests in which citizenship and residence differ. We built them to separate methods that read the wrong feature.

\begin{table*}[t]
\centering\footnotesize
\caption{Scenarios. ``Targets'' means the sender markets to or monitors people in that region.}\label{tab:scn}
\begin{tabular}{@{}lp{2.6cm}lp{5.0cm}p{3.9cm}@{}}
\toprule
Scenario & Sender host / targets & Receiver host & Residence mix & Purpose \\
\midrule
global mixed & OTHER / all & OTHER & EU .30, CA .25, IN .25, CN .05, OTHER .15 & residence alone decides the rule \\
EU resident, not targeted & IN / none & IN & EU .5, IN .5 & GDPR should not bind; DPDP should \\
India targets EU & IN / EU & CA & EU .7, IN .3 & GDPR (targeting) and DPDP both bind \\
EU to California & EU (established) / none & CA & EU .4, CA .4, IN .2 & establishment criterion; CCPA for CA \\
strict CN & CN / EU, CA & OTHER & CN .5, EU .2, CA .3 & stylised stricter regime \\
\midrule
resident, not citizen & OTHER / EU & OTHER & EU .8, OTHER .2 (citizens of IN, CN, OTHER) & non-Europeans living in the EU \\
citizen abroad, untargeted & OTHER / none & OTHER & OTHER .6, CA .4 (citizens mostly EU) & EU citizens, no EU rule binds \\
diaspora swap & OTHER / all & OTHER & EU .35, CA .30, IN .20, OTHER .15 (15\% citizens) & citizenship mostly wrong \\
\bottomrule
\end{tabular}
\end{table*}

\paragraph{Methods.}
\textit{Aggregate-only} shares only aggregates. \textit{Static (host region)} applies the rule of the sender's host region to every record.
\textit{Citizenship rule} treats the record as if residence equalled citizenship, and reads citizenship without noise.
\textit{Random feasible} uses the same belief and feasible set as \LCNB{} but picks a random individually rational bundle, with no bargaining.
\textit{\LCNB{}} is Algorithm~\ref{alg:lcnb}. \textit{Oracle} is \LCNB{} with true tags, versions and capacities, and $\epsilon=0$.

\paragraph{Metrics.}
\emph{Violation rate}: records violating some active rule divided by records shared.
\emph{Welfare}: realised $u_S+u_R$ with true attributes, minus unlock and verification costs.
\emph{Regret}: oracle welfare minus method welfare on the same stream.
\emph{Non-exempt share}: records shared with $T<3$ divided by records shared.
\emph{Diagnosis F1}: precision and recall of believed-active versus truly active (record, rule) pairs.

\paragraph{Two economic regimes.}
The \emph{default regime} (Table~\ref{tab:par}) charges leakage heavily. The \emph{binding regime} lowers the leakage weights ($\lambda_S=0.1$, $\lambda_R=0.05$), makes unlocking cheap and sets $\epsilon=0.1$.
We report both because, as Section~\ref{sec:default} shows, the default regime makes the rules rarely bind.

\paragraph{Protocol and scale.}
Each cell uses 5 seeds and 60 ticks (300 for drift), with 10 records per tick. The $\pm$ values are 95\% $t$-intervals over runs, pooled across grid cells and seeds, so they describe spread, not independent replication.
These are reduced-scale runs. Larger runs are configured in the repository.

\begin{table}[t]
\centering\small
\caption{Default parameters (abstract units).}\label{tab:par}
\begin{tabular}{@{}p{4.3cm}p{3.6cm}@{}}
\toprule
Parameter & Value \\
\midrule
Value to sender / receiver $v_S,v_R$ & 0.4 / 1.0 per record \\
Information value $I(T)$ & 1.0, 0.9, 0.6, 0.35 \\
Leakage weights $\lambda_S,\lambda_R$ & 0.5, 0.25 \\
Leakage proxy $\ell(T)$ & 1.0, 0.7, 0.25, 0.05 \\
Sensitive leakage multiplier & 3 \\
Fines $f$ (GDPR, DPDP, CCPA, PIPL) & 1.2, 1.0, 0.6, 1.5 \\
Sensitivity penalty multiplier $\kappa$ & 2 \\
Audit rate $p_a$; penalty share $\sigma$ & 0.3; 0.6 \\
Tolerance $\epsilon$ & 0.02 \\
Unlock: consent & 0.15 per record, $q$=0.7 \\
Unlock: mechanism & 2.0, $q$=0.6 \\
Verification cost $c_v$ & 0.2 \\
Tags: missing / wrong & 0.05 / 0--0.2 \\
Sensitive / child / opted-out share & 0.25 / 0.03 / 0.25 \\
\bottomrule
\end{tabular}
\end{table}

\section{Results}\label{sec:results}
\subsection{Default regime: low violations, but mostly aggregates}\label{sec:default}
Table~\ref{tab:main} gives the main comparison (5 natural scenarios, $p_w\in\{0,0.1,0.2\}$).
\LCNB{} has no violations and a regret of 0.05, against 0.65 for random feasible bundles and 0.75 for aggregate-only sharing.
The static host-region rule violates in 13.7\% of shared records, and the citizenship rule in 1.1\%.

\begin{table*}[t]
\centering\small
\caption{Default regime, five natural scenarios, three noise levels. Mean $\pm$ 95\% interval.}\label{tab:main}
\begin{tabular}{@{}lccccc@{}}
\toprule
Method & Violation rate & Welfare & Regret & Non-exempt share & Diagnosis F1 \\
\midrule
Aggregate-only & 0.000 $\pm$ 0.000 & 3.632 $\pm$ 0.002 & 0.747 $\pm$ 0.020 & 0.000 $\pm$ 0.000 & -- \\
Static (host region) & 0.137 $\pm$ 0.054 & 4.126 $\pm$ 0.067 & 0.254 $\pm$ 0.083 & 0.167 $\pm$ 0.063 & 0.844 $\pm$ 0.025 \\
Citizenship rule & 0.011 $\pm$ 0.004 & 4.330 $\pm$ 0.017 & 0.050 $\pm$ 0.013 & 0.041 $\pm$ 0.011 & 0.973 $\pm$ 0.006 \\
Random feasible & 0.000 $\pm$ 0.000 & 3.730 $\pm$ 0.013 & 0.650 $\pm$ 0.017 & 0.009 $\pm$ 0.006 & 0.957 $\pm$ 0.014 \\
LCNB & 0.000 $\pm$ 0.000 & 4.330 $\pm$ 0.014 & 0.050 $\pm$ 0.017 & 0.012 $\pm$ 0.007 & 0.957 $\pm$ 0.014 \\
Oracle & 0.000 $\pm$ 0.000 & 4.380 $\pm$ 0.020 & 0.000 $\pm$ 0.000 & 0.036 $\pm$ 0.011 & 1.000 $\pm$ 0.000 \\
\bottomrule

\end{tabular}
\end{table*}

\textbf{A caution.} The non-exempt share explains much of this. \LCNB{} shares 1.2\% of records in a non-exempt form, and the oracle 3.6\%.
Under these economics nearly everything is best sent as an aggregate, and aggregates are exempt by construction. Low violations are then easy.
Per-scenario violations (Table~\ref{tab:mainscn}) show that the static rule fails mostly in the global mixed scenario, where the host region carries no rule.
The citizenship rule does well because citizenship and residence agree for 90\% of records in these scenarios, and because it reads citizenship without noise.
These results do not show that \LCNB{} beats the citizenship rule in natural cases.

\begin{table}[t]
\centering\small
\caption{Violation rate by scenario (default regime).}\label{tab:mainscn}
\begin{tabular}{@{}lccc@{}}
\toprule
Scenario & Static & Citizenship & \LCNB{} \\
\midrule
global mixed & 0.604 & 0.018 & 0.000 \\
EU resident, not targeted & 0.031 & 0.031 & 0.000 \\
India targets EU & 0.052 & 0.009 & 0.000 \\
EU to California & 0.000 & 0.000 & 0.000 \\
strict CN & 0.000 & 0.000 & 0.000 \\
\bottomrule

\end{tabular}
\end{table}

\subsection{Residence versus citizenship}
Table~\ref{tab:contrast} uses the three stress scenarios, with audit rates 0.3 and 0.6 and sensitive shares 0.2 and 0.5.
Where citizenship and residence differ, the citizenship rule violates in 37\% of shared records for non-Europeans living in the EU, and in 21\% for EU citizens abroad with no EU rule binding.
Its diagnosis F1 is 0.26 in the second case. In the first, it predicts no active rule at all, so F1 is undefined.
\LCNB{} stays at or below 0.3\% and has F1 of 0.93 or higher.
The diaspora swap gives a smaller gap in violations (5.7\% against 0) even though citizenship is mostly wrong, because the wrongly chosen rules often demand similar safeguards.
The errors run in both directions. In the second case the citizenship rule applies a rule that does not bind (GDPR) and misses one that does (CCPA).

\begin{table}[t]
\centering\small
\caption{Stress scenarios, default regime. Regret is against the oracle.}\label{tab:contrast}
\resizebox{\columnwidth}{!}{%
\begin{tabular}{@{}llccc@{}}
\toprule
Scenario & Method & Violation & F1 & Regret \\
\midrule
resident, not citizen & Static (host region) & 0.527 & -- & 2.13 \\
 & Citizenship rule & 0.368 & -- & 1.56 \\
 & LCNB & 0.000 & 0.98 & 0.23 \\
\midrule
citizen abroad, untargeted & Static (host region) & 0.263 & -- & 0.35 \\
 & Citizenship rule & 0.213 & 0.26 & 0.44 \\
 & LCNB & 0.003 & 0.94 & 0.17 \\
\midrule
diaspora swap & Static (host region) & 0.552 & -- & 1.55 \\
 & Citizenship rule & 0.057 & 0.33 & 0.31 \\
 & LCNB & 0.000 & 0.93 & 0.21 \\
\bottomrule

\end{tabular}}
\end{table}

\subsection{Binding regime}
\begin{figure}[t]
\centering
\includegraphics[width=\columnwidth]{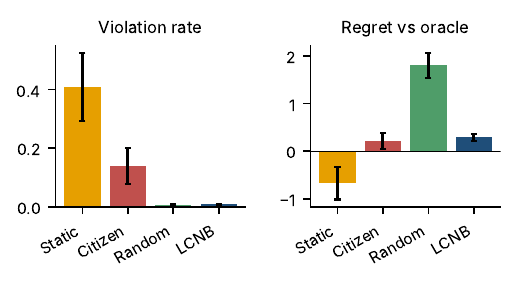}
\caption{Binding regime. Violation rate and regret by method (95\% intervals). Negative regret means higher expected welfare than the oracle, reached by violating (Section~\ref{sec:enf}).}
\label{fig:binding}
\end{figure}

In the binding regime (Table~\ref{tab:binding}, Fig.~\ref{fig:binding}) more records are shared in non-exempt form: 26\% for \LCNB{}, 32\% for the oracle.
\LCNB{} violates in 0.9\% of shared records. The tolerance $\epsilon=0.1$ bounds the violation probability of each group under the model, and the observed rate is far below it.
The citizenship rule violates in 13.9\% and the static rule in 40.8\%. Table~\ref{tab:bindscn} gives the values by scenario.

Welfare ranks the methods differently. The static rule has welfare 7.10 and the citizenship rule 6.23, against 6.14 for \LCNB{} and 6.43 for the oracle.
Both rules beat the oracle on this measure at audit rate $0.3$, because the expected fine is small compared with the gain from sharing more.
This is the situation Proposition~\ref{prop:enf} describes. Welfare here counts fines only, so it should not be read as a recommendation to violate. Compliance is a constraint in \LCNB{}, not a term to trade off.

\begin{table}[t]
\centering\small
\caption{Binding regime, eight scenarios, audit rate 0.3. Means; intervals are in Fig.~\ref{fig:binding} for violation and regret.}\label{tab:binding}
\footnotesize\setlength{\tabcolsep}{3pt}
\begin{tabular}{@{}lccccc@{}}
\toprule
Method & Viol. & Welfare & Regret & Non-ex. & F1 \\
\midrule
Static (host region) & 0.41 & 7.10 & -0.67 & 0.61 & 0.85 \\
Citizenship rule & 0.14 & 6.23 & 0.21 & 0.33 & 0.78 \\
Random feasible & 0.01 & 4.63 & 1.81 & 0.18 & 0.97 \\
LCNB & 0.01 & 6.14 & 0.29 & 0.26 & 0.97 \\
Oracle & 0.00 & 6.43 & 0.00 & 0.32 & 1.00 \\
\bottomrule

\end{tabular}
\end{table}

\begin{table*}[t]
\centering\small
\caption{Binding regime by scenario: violation rate and welfare.}\label{tab:bindscn}
\begin{tabular}{@{}lcccccc@{}}
\toprule
 & \multicolumn{3}{c}{Violation rate} & \multicolumn{3}{c}{Welfare} \\
\cmidrule(lr){2-4}\cmidrule(lr){5-7}
Scenario & Static & Citizenship & \LCNB{} & Citizenship & \LCNB{} & Oracle \\
\midrule
global mixed & 0.848 & 0.034 & 0.019 & 6.26 & 6.34 & 6.69 \\
EU resident, not targeted & 0.043 & 0.044 & 0.010 & 6.64 & 7.10 & 7.50 \\
India targets EU & 0.333 & 0.025 & 0.008 & 5.28 & 5.38 & 5.60 \\
EU to California & 0.000 & 0.000 & 0.000 & 4.79 & 4.79 & 4.79 \\
strict CN & 0.000 & 0.000 & 0.000 & 4.79 & 4.79 & 4.79 \\
diaspora swap & 0.857 & 0.139 & 0.015 & 5.52 & 6.10 & 6.43 \\
resident, not citizen & 0.794 & 0.563 & 0.011 & 7.05 & 6.06 & 6.30 \\
citizen abroad, untargeted & 0.389 & 0.308 & 0.006 & 9.49 & 8.58 & 9.37 \\
\bottomrule

\end{tabular}
\end{table*}

\subsection{Enforcement}\label{sec:enf}
\begin{figure}[t]
\centering
\includegraphics[width=\columnwidth]{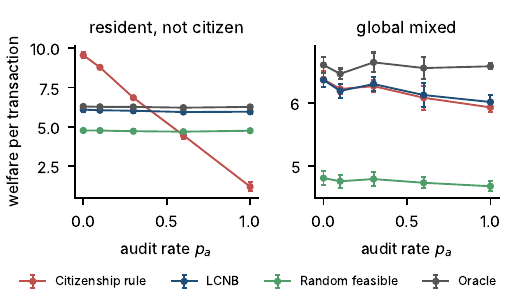}
\caption{Welfare against audit rate in the binding regime. A method that violates gains when audits are rare and loses when they are frequent.}
\label{fig:enforce}
\end{figure}

Figure~\ref{fig:enforce} varies the audit rate in the binding regime.
For non-Europeans in the EU, the citizenship rule has welfare 9.56 at $p_a=0$, 6.85 at $0.3$ and 4.46 at $0.6$, against 6.2--6.3 for the oracle at each rate.
Linear interpolation puts the crossing with the oracle at $p_a\approx0.37$, which is the empirical counterpart of $p^*$ in Proposition~\ref{prop:enf}. It is an interpolation between two measured points, not a fitted value.
\LCNB{} welfare barely moves with $p_a$ (6.09 to 5.96), and its violation rate stays at 0.5--1.1\%. Its choices depend on the tolerance $\epsilon$, not on $p_a$.
A Stackelberg enforcer that picks $p_a$ to minimise harm plus audit cost therefore has little to learn from \LCNB{}. It matters only for agents that violate because they are ignorant of the rule.
In the global mixed scenario the citizenship rule and \LCNB{} stay within 0.1 welfare of each other at every audit rate.

\subsection{Violation tolerance}
\begin{figure}[t]
\centering
\includegraphics[width=0.92\columnwidth]{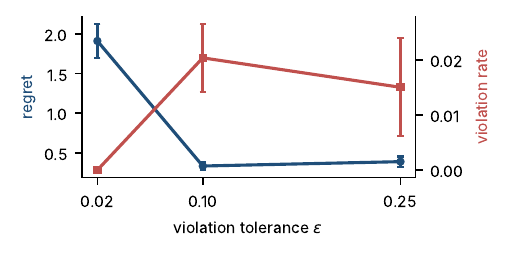}
\caption{Regret and violation rate of \LCNB{} against $\epsilon$.}
\label{fig:eps}
\end{figure}

Table~\ref{tab:eps} and Fig.~\ref{fig:eps} vary $\epsilon$ in the global mixed scenario with cheap safeguards.
At $\epsilon=0.02$ \LCNB{} shares no non-exempt bundle, and its regret is 1.91. At $\epsilon=0.1$ regret falls to 0.33 and violations rise to 2.0\% of shared records.
At $\epsilon=0.25$ regret is 0.39 and violations 1.5\%, so there is no gain beyond about 0.1 in this setting.
This matches Proposition~\ref{prop:noise}. The 5\% of records with a missing residence tag have a posterior equal to the prior, with EU mass 0.30, and no non-exempt bundle passes $\epsilon=0.02$.
The observed violation rate (2.0\%) is below the nominal 10\% because $\epsilon$ bounds each group's violation probability under the model, and most groups have far lower probability.
This is the central design choice of the method. In our runs it mattered more than the choice of bargaining rule.

\begin{table}[t]
\centering\small
\caption{Violation tolerance (global mixed, cheap safeguards, $p_m=p_w=0.05$).}\label{tab:eps}
\resizebox{\columnwidth}{!}{%
\begin{tabular}{@{}rlccccr@{}}
\toprule
$\epsilon$ & Method & Welfare & Regret & Violation & Non-exempt & Unlocks \\
\midrule
0.02 & LCNB & 4.79 & 1.91 & 0.000 & 0.00 & 0 \\
 & LCNB, no unlock & 4.79 & 1.91 & 0.000 & 0.00 & 0 \\
 & LCNB, no verify & 4.79 & 1.91 & 0.000 & 0.00 & 0 \\
 & Oracle & 6.70 & 0.00 & 0.000 & 0.36 & 26 \\
\midrule
0.1 & LCNB & 6.27 & 0.33 & 0.020 & 0.29 & 30 \\
 & LCNB, no unlock & 6.14 & 0.47 & 0.019 & 0.25 & 0 \\
 & LCNB, no verify & 6.28 & 0.32 & 0.020 & 0.29 & 30 \\
 & Oracle & 6.61 & 0.00 & 0.000 & 0.35 & 25 \\
\midrule
0.25 & LCNB & 6.14 & 0.39 & 0.015 & 0.26 & 32 \\
 & LCNB, no unlock & 6.00 & 0.53 & 0.013 & 0.22 & 0 \\
 & LCNB, no verify & 6.15 & 0.38 & 0.015 & 0.26 & 32 \\
 & Oracle & 6.52 & 0.00 & 0.000 & 0.34 & 29 \\
\bottomrule

\end{tabular}}
\end{table}

\textbf{Unlock and verify.} Unlocking helps a little once $\epsilon\ge0.1$: welfare 6.27 against 6.14 without it. Under the default economics it never fires.
Verification \emph{never fired} in any run (zero verification calls across all configurations in this paper).
The intended case needs a bundle whose mechanism level exceeds the pair's true capacity and a sender that would otherwise rely on the claim. In our scenarios, the bundles chosen did not need it.
We therefore have no evidence that verification helps, and we report it as untested.

\subsection{Law drift}\label{sec:drift}
\begin{figure}[t]
\centering
\includegraphics[width=\columnwidth]{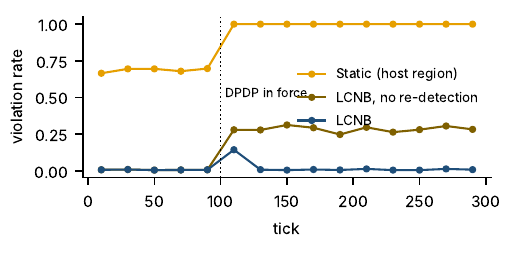}
\caption{Violation rate by 20-tick window in the binding regime. DPDP commences at tick 100.}
\label{fig:drift}
\end{figure}

In the drift test the DPDP's operative provisions switch on at tick 100.
In the default regime the agent without re-detection and the agent with it behave the same: zero violations and the same welfare, because both share aggregates.
Diagnosis recall differs (0.50 without re-detection, 0.97 with it), so the diagnosis error exists but has no cost there.
In the binding regime (Fig.~\ref{fig:drift}, Table~\ref{tab:drift}) it does. After commencement the agent without re-detection violates in 28\% of shared records. \LCNB{} violates in 28\% during its 10-tick lag and in 0.4\% afterwards, which is 14\% over the first 20 ticks.
This matches Proposition~\ref{prop:drift} with $q\approx0.28$.
Without re-detection welfare is slightly higher (5.97 against 5.81), for the same reason as in Section~\ref{sec:enf}: it shares more and pays expected fines.

\begin{table}[t]
\centering\small
\caption{Law drift: overall violation rate, diagnosis recall and regret.}\label{tab:drift}
\resizebox{\columnwidth}{!}{%
\begin{tabular}{@{}llccc@{}}
\toprule
Regime & Method & Violation & Recall & Regret \\
\midrule
default economics & LCNB & 0.000 & 0.97 & 0.18 \\
 & LCNB, no re-detection & 0.000 & 0.50 & 0.18 \\
 & Static (host region) & 0.700 & 0.00 & 2.42 \\
\midrule
binding regime & LCNB & 0.016 & 0.97 & 0.22 \\
 & LCNB, no re-detection & 0.191 & 0.50 & 0.07 \\
 & Static (host region) & 0.895 & 0.00 & 0.10 \\
\bottomrule

\end{tabular}}
\end{table}

\subsection{Which tags matter}
Table~\ref{tab:feat} removes one tag at a time from the diagnosis (default regime, global mixed and India targets EU).
Dropping the residence tag raises regret from 0.064 to 0.121 and cuts diagnosis recall from 0.905 to 0.500. Dropping the sensitive flag raises regret to 0.122.
Dropping consent evidence or the opt-out flag changes nothing in this regime, and the child and legitimate-use flags change regret by under 0.01.
Using the observed tag as if it were true (no posterior) raises violations to 1.4\%, against zero for the full posterior.
At $p_w=0.2$ dropping residence leaves regret unchanged (0.115) and gives a higher F1 (0.92 against 0.87). We did not investigate why. A likely cause is that a 0.5 threshold on a noisy posterior makes more wrong calls than the prior does.

\begin{table}[t]
\centering\small
\caption{Feature ablation (default regime). Tag removed from the diagnosis.}\label{tab:feat}
\resizebox{\columnwidth}{!}{%
\begin{tabular}{@{}lccc@{}}
\toprule
Tag removed & Regret & Recall & Violation \\
\midrule
None (full LCNB) & 0.064 $\pm$ 0.041 & 0.905 & 0.000 \\
Plug-in tags (no posterior) & 0.076 $\pm$ 0.039 & 0.910 & 0.014 \\
Residence & 0.121 $\pm$ 0.048 & 0.500 & 0.000 \\
Sensitive flag & 0.122 $\pm$ 0.039 & 0.905 & 0.000 \\
Consent evidence & 0.064 $\pm$ 0.041 & 0.905 & 0.000 \\
Opt-out & 0.064 $\pm$ 0.041 & 0.905 & 0.000 \\
Child flag & 0.072 $\pm$ 0.039 & 0.905 & 0.000 \\
Legitimate use & 0.072 $\pm$ 0.039 & 0.905 & 0.000 \\
\bottomrule

\end{tabular}}
\end{table}

\subsection{Bargaining solutions}
Table~\ref{tab:barg} compares solutions as the sender's value $v_S$ varies (0.2, 0.4, 1.0; the table pools them).
The Monotonic Concession Protocol gives the same welfare as the Nash solution in every cell, as Proposition~\ref{prop:nash} predicts, and takes 11--26 rounds per tick, depending on $v_S$.
Kalai--Smorodinsky is slightly lower in all three cells. The utilitarian rule has the highest total, by construction, since welfare is the sum. The Rubinstein approximation is also above Nash on the sum.
The experiments do not measure how surplus is split between sender and receiver, and that is where these solutions differ most. We leave it for future work.
All solutions have zero violations here, but in this default regime almost everything shared is an aggregate, so this says little about compliance. We did not repeat the comparison in the binding regime.

\begin{table}[t]
\centering\small
\caption{Bargaining solutions (global mixed; pooled over $v_S$, so the intervals mostly reflect differences between cells).}\label{tab:barg}
\begin{tabular}{@{}lccc@{}}
\toprule
Solution & Welfare & Regret & Rounds \\
\midrule
LCNB & 4.78 $\pm$ 0.72 & 0.590 & 0.0 \\
LCNB, MCP (Zeuthen) & 4.78 $\pm$ 0.72 & 0.590 & 17.5 \\
LCNB, Kalai-Smorodinsky & 4.73 $\pm$ 0.74 & 0.635 & 0.0 \\
LCNB, Rubinstein & 4.83 $\pm$ 0.68 & 0.537 & 19.1 \\
LCNB, utilitarian & 4.85 $\pm$ 0.69 & 0.519 & 0.0 \\
Random feasible & 3.69 $\pm$ 0.23 & 1.684 & 0.0 \\
\bottomrule

\end{tabular}
\end{table}

\textbf{Leakage proxy.} We calibrated the leakage proxy with a nearest-neighbour linking attack on synthetic records (Appendix~C).
The measured re-identification rates for the four transformation levels are 1.00, 0.71, 0.06 and 0.02, against our default proxy of 1.00, 0.70, 0.25 and 0.05.
The proxy is close for pseudonymised data and conservative for the other two. With the calibrated values \LCNB{} still has zero violations.

\section{Discussion}
\paragraph{What the results support.}
(i) When residence and citizenship differ, rules based on citizenship or on host region fail in both directions. A belief over the true attributes avoids this. (ii) The tolerance $\epsilon$ controls a clear trade-off between caution and utility, and Proposition~\ref{prop:noise} explains its shape.
(iii) Re-reading rule versions matters once a new rule changes what is lawful, and Proposition~\ref{prop:drift} predicts the size of the spike.
(iv) The Monotonic Concession Protocol reproduces the Nash solution. We saw no compliance difference between bargaining rules, but only in the default regime.

\paragraph{What they do not support.}
(i) In the default regime, \LCNB{} shares almost only aggregates. Its zero violations there are weak evidence. (ii) The citizenship rule is a strong baseline in natural scenarios, partly because we gave it noise-free citizenship.
(iii) In our welfare measure a violating method can beat the oracle at low audit rates. We report this and do not call it a result in favour of violating. (iv) Verification was never used, and unlock only helped at higher tolerance.
(v) The enforcement layer is not a full Stackelberg solution. We swept $p_a$ and read off a crossing.

\paragraph{Threats to validity.}
The data are synthetic. The rules are stylised and several clauses come from our own knowledge rather than a legal review. Fines and values are abstract.
There is one sender--receiver pair, and a mediator computes the Pareto frontier. Runs are short (5 seeds, 60 ticks). Parameters such as $\epsilon$ and $\lambda$ were chosen by us, and conclusions may move with them. Grid-cell pooling widens the intervals but does not give independent replicates.

\paragraph{Relation to the wider transaction problem.}
Regime uncertainty is one source of constraint. Ownership, purpose limits, internal policy, intellectual-property and confidentiality screening, trust, and incentives also shape which flows are allowed and useful.
In our model they enter as extra dimensions in the bundle, extra terms in the utilities, and a reputation. Extending the constraint compiler to those sources is the main next step.

\section{Conclusion}
We framed legally uncertain data transactions between ML agents as a game with a constraint structure, a belief over which rules apply, optional safeguards and verification, and a bargaining step. The algorithm, \LCNB{}, is simple to state and to run.
In tests it avoids the mistakes of citizenship and host-region rules when residence and citizenship differ, and it handles rule changes with a short lag.
The tolerance $\epsilon$ is the main lever. The experiments also showed what is not yet established: the benefit of verification, the split of surplus, and results at larger scale.
All code, configurations and the scripts that build every table and figure are released.

\appendix
\section{Legal encoding (summary)}
Rules are research abstractions of public texts. \textbf{GDPR} (Reg.\ 2016/679): scope by establishment (Art.~3(1)) or by targeting EU residents (Art.~3(2)); purpose and retention limits (Art.~5); basis, child and special-category consent (Art.~6, 8, 9); processor contract (Art.~28); transfer safeguards (Art.~44--49).
\textbf{DPDP Act 2023}: processing in India or offering goods or services to people in India (s.~3); consent and legitimate uses (s.~6, 7); processor contract (s.~8); children (s.~9); restriction by notified destinations (s.~16); operative provisions not in force before commencement.
\textbf{CCPA/CPRA}: California residents and covered businesses; sale or sharing to third parties triggers notice, opt-out and a contract (\S1798.100, .120, .135, .140); service-provider terms; limits on sensitive data (\S1798.121).
\textbf{PIPL-like}: a stylised stricter regime with separate consent and a security assessment for cross-border transfer; it is not a faithful encoding.
The repository file \texttt{docs/LEGAL\_ENCODING.md} marks each row as checked in public sources or taken from our own knowledge. Citizenship is read by no rule.

\section{Reproduction}
\small
\begin{verbatim}
pip install -e . pytest && pytest -q
python -m eregime.run \
  --config paper/configs/p1_main.yaml \
  --out paper/results/p1_main
python paper/make_assets.py
cd paper && pdflatex main.tex
\end{verbatim}
\normalsize
Configurations \texttt{p1}--\texttt{p11} in \texttt{paper/configs} reproduce every number here. Larger grids are in \texttt{configs/}.

\section{Leakage calibration}
Level 0 (raw) is treated as fully identifying. Level 1 (pseudonymised) keeps features; the attacker holds a noisy auxiliary copy and links by nearest neighbour.
Level 2 generalises each feature to four bins and adds Laplace noise. Level 3 releases means of random buckets of ten, and success means the nearest auxiliary record lies in the bucket.
This is one attack on synthetic data. It is a calibration aid, not a privacy guarantee.

\end{document}